\documentclass[twoside,twocolumn]{article}

\usepackage{graphicx}
\usepackage{blindtext} 
\usepackage{amsmath}
\usepackage[sc]{mathpazo} 
\usepackage[T1]{fontenc} 
\usepackage{microtype} 

\usepackage[english]{babel} 

\usepackage[hmarginratio=1:1,top=32mm,columnsep=20pt]{geometry} 
\usepackage[hang, small,labelfont=bf,up,textfont=it,up]{caption} 
\usepackage{booktabs} 

\usepackage{lettrine} 

\usepackage{enumitem} 
\setlist[itemize]{noitemsep} 

\usepackage{abstract} 

\usepackage{titlesec} 
\renewcommand\thesection{\Roman{section}} 
\renewcommand\thesubsection{\roman{subsection}} 
\titleformat{\section}[block]{\large\scshape\centering}{\thesection.}{1em}{} 
\titleformat{\subsection}[block]{\large}{\thesubsection.}{1em}{} 

\usepackage{titling} 

\usepackage{hyperref} 

\pretitle{\begin{center}\Huge\bfseries} 
\posttitle{\end{center}} 
\title{$\Psi$DM and the Bullet Cluster: Couple Inflaton to Higgs Field} 
\author{%
\textsc{Chia-Li Hsieh}\\[1ex]
\normalsize Department of Physics, National Taiwan University, Taipe 106 Taiwan \\ 
}
\date{} 
\renewcommand{\maketitlehookd}{%
\begin{abstract}
Electroweak theory is extended to $SU_L(2)\times U_Y(1)\times U_{\phi}(1)$, where $U_{\phi}(1)$ comes from the coupling of inflaton field to Higgs field. There arises an extremely light boson, $Y_{\mu}$. We resemble it as the wavelike dark matter, $\Psi$DM, through the dynamical analysis. The interactions of $Y_{\mu}$ may provide an explanation for the Bullet Cluster observations. Besides, a possible mechanism to produce $Y_{\mu}$ is discussed.

\end{abstract}
}

\begin{document}

\maketitle


\section{Introduction}


Dark matter (DM) particles, excluded from the Standard Model (SM), abundantly exist in the Universe and form halos around galaxies [1]. DM is assumed to be collisionless, moving nonrelativistically and interacting primarily through gravity in the cold dark matter (CDM) model [2]. This model speaks well for the large scale structure of cosmos, which is shaped by DM. However, it suffers from the core-cusp problem [3], in which the prediction of CDM model is inconsistent with the observations at small scale. That is, an increasingly sharp density profile [4] is cast, whereas flat-core density profiles of DM dominated dwarf galaxies [5] and low surface bright galaxies [6] have been observed. Apparently, it seems necessary to come up with new physics concepts in DM formation. Two examples are given to complement the mechanism, $\Psi$DM [7] and the Bullet Cluster (BC) [8].  We will connect these two sectors with a concordant model.

\par

Firstly, to remedy the cuspy problem, an ultralight boson particle $\sim 10^{-22}$ eV, termed $\Psi$DM, is introduced [7] by countering the gravity through uncertainty principle of galactic wavelengths and forming clumps through Bose-Einstein condensation (BEC) by self-gravity. $\Psi$DM model has not got much attention until  recently the impressive work, simulated via adaptive mesh refinement and GPU technology by Schive et al. [9], has shown not only an obvious flat-core profile of DM distribution on sub-galactic scale but also an indistinguishable large scale structure of cosmic filaments and voids from CDM model. Consequently, it greatly enhances the confidence to adopt $\Psi$DM as a candidate of DM particle. In another scenario, galaxy cluster collision is the process of merging clumps of intracluster gases (emitting X-ray), dark matter (observed by gravitational lensing), and galaxies. Observing the typical case of BC [8], the dark-matter clumps pass through each other without a prominent interaction with each other, which may slow down the DM clumps. On the other hand, the study [10] indicates that a non-gravitational force of DM may interact with visible matter, the SM particles, by measuring the spatial offsets. This study [10] provides an evidence to testify whether the DM particle is collisionless or not.

\par
Despite the success of $\Psi$DM model, the existence of such an extremely light particle is implicit [11] and the mechanism of non-gravitational force in BC also lacks proper explanations [12]. We observe that an ultralight boson particle arises from the weakly coupling of inflation field with Higgs field, which will also provide the interpretation to BC naturally. As one can conceive, the inflation model may go further to be incorporated in SM and DM particle can be formulated in the extended SM.

\par

In the cosmic inflation regime [13], the inflation field, or inflaton, dubbed F here, exerts a great negative pressure and drives the very early universe to expand quickly through its plateau-like potential. This theory may be testified via Cosmic Background Explorer (COBE), Wilkinson Microwave Anisotropy Probe (WMAP) and Planck mission [14], since the inflation magnifies the small fluctuations into mega fluctuations, e.g. the large scale structure and anisotropies in CMB (e.g., [15]). Moreover, in the post-inflation era, the inflaton theoretically loses most of the energy and decays into other particles dramatically during reheating phase [16]. The core assumption we propose in this letter is that the inflatons may decay but not die out. After reheating, they become very dilute and evenly distributed in the Universe that particles are very rarely collided by inflatons and corresponds to an extremely small but non-negligible  coupling constant.

\par

Even so, how could the inflaton still affect the Universe? Could we incorporate it into SM without harming the perfect mechanism of SM? The inspiration is to weakly couple inflaton field with Higgs field, analogous to the cooling process of a ferromagnet under an external field. To mimic the process, since in SM the Higgs fields are set to be a doublet, $\Phi=\begin{pmatrix}
        \phi_{+} \\
        \phi_{0}
      \end{pmatrix}$ [17], we also arrange the inflation fields into the same form, $F=\begin{pmatrix}
     F_1 \\
     F_0
   \end{pmatrix}$. Then there arises a natural coupling with a constant c,

\begin{equation}
  L_1=c(F^{+}\Phi + \Phi^{+}F)
\end{equation}

where F and $\Phi$ play the role of the external field and magnetization, respectively.
\par

Moreover, in the Higgs-field spanned internal space, $L_1$ drives a local potential minimum. One could find that in this scenario a perturbation energy will stimulate an extra phase from the potential minimum. Consequently, a new gauge transformation $U_\phi(1)$ and a gauge field are introduced to preserve the gauge symmetry caused by the phase. A further application to coupling the new gauge field to inflaton implies the extremely light mass. Through the nonrelativistic approach, we resemble $Y_{\mu}$ as $\Psi$DM using the mechanism of Bose-Einstein Condensation (BEC) [18] and its extremely light mass. Additionally, in BC the bizarre behavior of DM of exerting the force on ordinary particles but ignoring self-interactions may be qualitatively explained by considering the interaction Lagrangian of $Y_{\mu}$ via mimicking Compton scattering and $\gamma-\gamma$ interaction separately. In the end, we talk about a possible mechanism to produce $Y_{\mu}$ abundantly, related to the coupling constants.


\section{Theory}


To begin with, we illustrate an old phenomenon: a ferromagnet is perturbed by an external field H. We may write down the free energy as [17]
$E_F=\int d^{3}x[\frac{1}{2}(\nabla S)^{2}+m(T-T_{c})S^{2}+nS^{4}-H \cdot S]$. Obviously, a local potential minimum has the preferred orientation dependence along H. Analogous to $H \cdot S$ , we have $  L_1=c(F^{+}\Phi + \Phi^{+}F)$ , as mentioned earlier. In polar-coordinate representation of field-space, set $\phi=\rho e^{i\theta}=\rho(\cos \theta +i \sin \theta)$, such that
$L_1=c[\begin{pmatrix}
         F^{*}_1 & F^{*}_0
       \end{pmatrix}\begin{pmatrix}
        \phi_{+} \\
        \phi_{0}
      \end{pmatrix}+\begin{pmatrix}
         \phi_{+}^{*} & \phi_{0}^{*}
       \end{pmatrix}\begin{pmatrix}
        F_{1} \\
        F_{0}
      \end{pmatrix}]=c(\cos\theta +b \sin\theta)$, where

$a=(F_1^{*}\rho_{+}+F_0^{*}\rho_{0}+\rho_{+}F_1 + \rho_{0}F_{0})$,

$b=i(F_1^{*}\rho_{+}+F_0^{*}\rho_{0}-\rho_{+}F_1 - \rho_{0}F_{0})$. The potential of Higgs field,
\begin{flalign}
  V&=\mu^2(\Phi^{*}\Phi)+\lambda(\Phi^{*}\Phi)^{2}-c(F^{*}\Phi+\Phi^{*}F) \nonumber\\
  &=\mu^2(\Phi^{*}\Phi)+\lambda(\Phi^{*}\Phi)^{2}-c(a\cos\theta+b\sin\theta)
\end{flalign}

, which has a potential minimum along $\theta$ in internal space spanned by $\Phi$ as shown in figure 1. Note that this minimum is independent of $\mu^2$.

\begin{figure}
  \centering
  \includegraphics[width=0.5\textwidth]{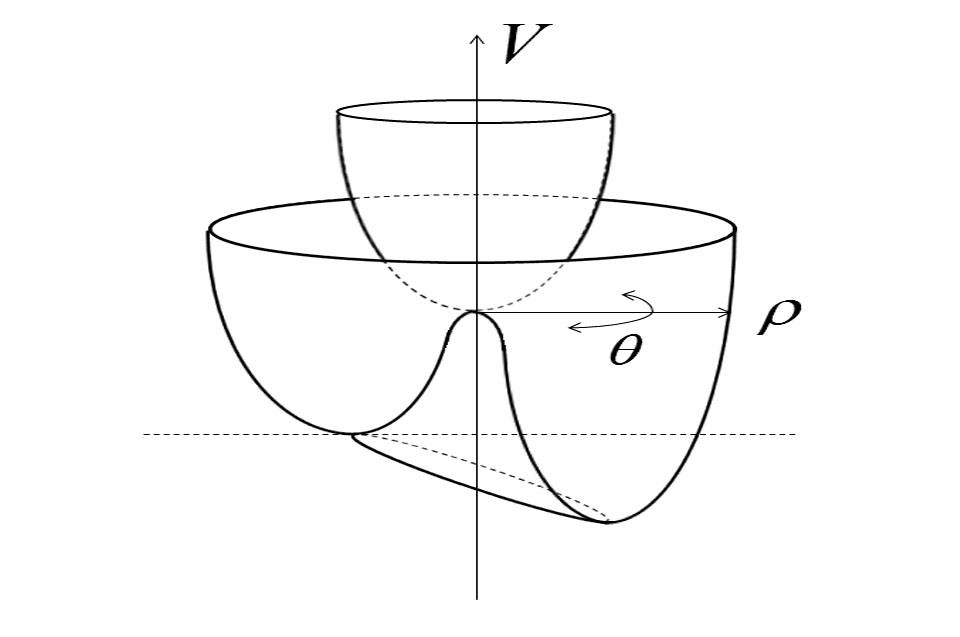}
  \caption{the tilted potential composed by affected Higgs fields shows a $\theta$  dependence}\label{f1}
\end{figure}

\begin{figure}
  \centering
  \includegraphics[width=0.5\textwidth]{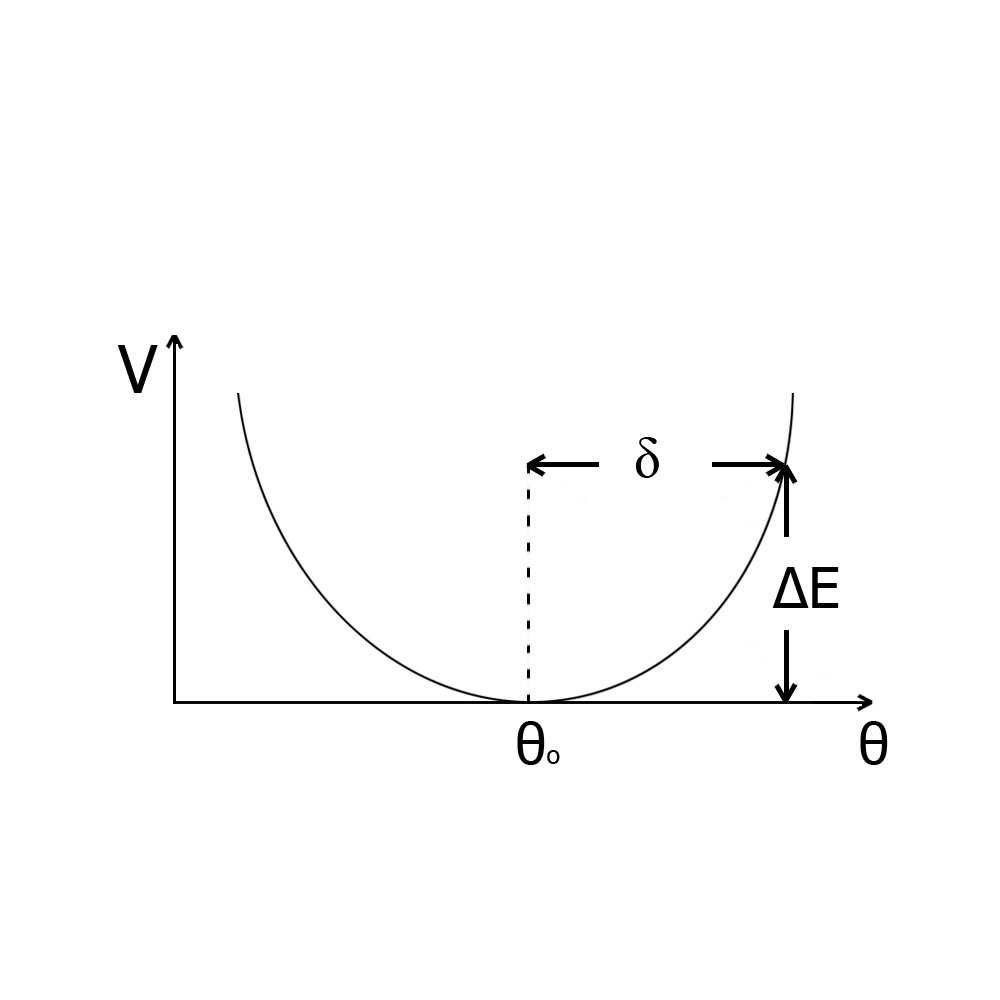}
  \caption{an excited energy $\Delta E$ stimulates $\delta$ from the potential minimum.} \label{f2}
\end{figure}

\par

In the valley plain of the potential minimum, we can loosely choose $\theta$ as a phase to be transformation invariant since the potential is rather flat in the vicinity of the minimum. One should note that $<\theta> =\theta_0$ corresponding to the potential minimum. However, in contrary to moving around $\theta_0$, the perturbation energy $\Delta E$ enters in the potential and excite an energy difference $\Delta V$ that will trigger a large displacement $\delta$ deviated from $\theta_0$, as shown in figure 2. Since the perturbation $\delta$ expressed in polar coordinate performs as the phase difference, we extend the phase $\delta$ to a transforming invariance and pursue a gauge symmetry. In the case of phase transformation of global gauge, $\Phi \rightarrow \Phi'=e^{i\delta \Phi}$ one can find that $V\rightarrow V'\approx V$, where we set $\theta=\theta_0+\delta$ and expand around $\theta_0$ in the potential of Eq. (2). It is $\frac{\partial V}{\partial \theta}|_{\theta_0}=0$ that the Lagrangian is approximately symmetry to the first order. Therefore, in addition to choosing the ordinary $U(1)=e^{i\theta}$ transformation, we can make an extra $U_\phi(1)=e^{i\delta}$  transformation to seek a global symmetry. Similarly, we have to introduce a new gauge field $C_{\mu}$  in the local gauge transformation. A covariant derivative $D_{\mu}\Phi=\partial_{\mu}\Phi+iC_{\mu}\Phi$ is introduced to derive $D_{\mu}\Phi\rightarrow D'_{\mu}\Phi'=e^{i\delta(x)}D_{\mu}\Phi$.  Also, the gauge field transforms as $C_{\mu}\rightarrow C'_{\mu}=C_{\mu}-\frac{\partial}{\partial x^{\mu}}\delta$.  As a result, in our semi-classical approach the $\delta$, aroused by $\Delta E$, in the coupling $L_1=c(F^{+}\Phi+\Phi^{+}F)$  forces us to pick up a transformation, termed $U_\phi(1)=e^{i\delta}$.

\par
For the sake of discussing the effect of $U_\phi(1)$ in Standard Model, the dynamics of $C_\mu$ and its coupling need elucidation. Since the $U_\phi(1)$ simply comes from the potential deviation in the internal space, on which the $U_Y(1)$ relies (as shown in Figure 1), $U_\phi(1)$ and $U_Y(1)$ should share the same weakhyper charge, Y, but differ in strength. Accordingly, we denote the coupling constant of $U_\phi(1)$ as $g^{''}$ to be distinguished from $g'$ of $U_Y(1)$. Thus, we are coming to face the scenario of a modified electroweak theory with $SU_L(2)\times U_Y(1)\times U_{\phi}(1)$. As for the total covariant derivative of Higgs field, we have $D_{\mu}\Phi=(\partial_{\mu}+i\frac{1}{2}g\tau \cdot A_{\mu}+i\frac{Y}{2}g'B_{\mu}+i\frac{Y}{2}g^{''}C_{\mu})\Phi$ and $L_2=(D_{\mu}\Phi)^{+}(D^{\mu}\Phi)$.

\par
Focusing on the Lagrangian of massive terms of $L_2$, we have

\begin{multline}
    L'_2 =M_{w}^{2}W^{+}_{\mu}W^{\mu-} + \nonumber \\
\frac{v^{2}}{8} \begin{pmatrix}
                   A^{3}_{\mu} & B_{\mu} & C_{\mu}\end{pmatrix} \begin{pmatrix}
                                g^2 & -gg' & -gg^{"} \\
                                -gg' & g'^{2} & g'g^{"} \\
                                -gg^{"} & g^{'}g^{"} & g^{"2}
                              \end{pmatrix} \begin{pmatrix}
                                             A^{\mu3} \\
                                             B^{\mu} \\
                                             C^{\mu}
                                           \end{pmatrix}
\end{multline}

Orthogonal diagonalization of the matrix leads to a massive  $Z_{\mu}=\frac{1}{\sqrt{g^2+g'^2+g^{"2}}}(gA^{3}_{\mu}-g'B_{\mu}-g^{"}C_{\mu})$,
a massless photon $=\frac{1}{\sqrt{g^{2}+g'^{2}}}(g'A^{3}_{\mu}+gB_{\mu})$
and a new massless field,
$Y_{\mu}=\frac{1}{\sqrt{g^{2}g^{"2}+g'^{2}g^{"2}+(g^{2}+g'^{2})^2}}(gg^{"}A^{3}_{\mu}-g'g^{"}B_{\mu}+(g^{2}+g'^{2})C_{\mu})$.

To avoid harming the perfection of electroweak theory, $g^{''}$  would be very small. Correspondingly, $Z_{\mu}$  remains intact and we have a neat solution, $Y_{\mu}=C_{\mu}$. From the gauge transformation of $U_\phi(1)$, we should know that $Y_{\mu}$ can only be interacted via Y charge.

\par

Furthermore, one may consider another regime that the U(1) symmetry is broken by inflatons. The case of symmetry breaking of inflaton coupling to the gauge field via electric charge has been studied [19], in which they infer as the origin of the primordial magnetic field [20]. From the coupling, $L_1=c(F^{+}\Phi + \Phi^{+}F)$, the inflaton   carrying the weakhyper charge, Y=1, is apparent. Under the circumstances of connecting inflaton $F_0$ with massless field $Y_{\mu}$ we couple them by U(1) symmetry via Y=1 and a coupling constant $g^{'''}$. Gauge symmetry brings us

\begin{flalign}
  L_{3}&=(D_{\mu}F_{0})^{\dagger}(D^{\mu}F_{0})\nonumber \\
  &=(D_{\mu}F_{0})^{2}+g^{'''2}Y_{\mu}Y^{\mu}F^{2}_{0}.
\end{flalign}

After inflation, the inflation field rolls into the potential minimum. In this regime, it shifts to have $<F_0>=V_0 \sim 10^{-4}M_{pl} $[19]. Indeed, $F_0$ plays the role of Higgs field and symmetry breaking brings

\begin{flalign}
  L_{mass} & =\frac{1}{2}2g^{'''2}V_{0}^{2}Y_{\mu}Y^{\mu}=\frac{1}{2}m^{2}Y_{\mu}Y^{\mu}
\end{flalign}

In the early universe, inflaton F evolves into the reheating phase [16], when it passes most of its energy to other particles and decays dramatically after inflation. Far away from the phase, we postulate that inflation fields are not extinct but decay to be extremely scarce. Furthermore, from the particle physics perspective of giving mass by Higgs field, the particle that collides with Higgs boson has been conferred the mass. The more frequently the particle hits, the heavier mass it carries. Conceptually, comparing to a heavier particle, a lighter particle would have a small coupling constant and then correspond to a longer mean free path, as shown in Figure 3. In our case, since the density of inflaton is very dilute in space,  a particle which seldom collides with inflatons needs a very long mean free path. Therefore, it effectively reflects a very small coupling constant. From Eq (3), after symmetry breaking we may expect a very small $g^{'''}$, since the density of F is extremely low in our present universe. In another way, in $L_1=c(F^+\Phi+\Phi^+F)$ , c represents the chance for converting between $\Phi$ and F. In the low energy limit, the interaction between $\Phi$ and F is rather faint by the fact that F is so dilute. This corresponds to an extremely small c. From figure 2, we see that the tiny perturbative energy entering the potential can trigger the $Y_{\mu}$ particle, or reciprocally we may assume that the energy $\Delta E$ converts to $Y_{\mu}$ and expect that $g^{'''}\propto c$[21]. Given the extremely light mass and dynamics of $Y_{\mu}$ to what is this field corresponding? We present two applications,$\Psi$DM[7] and Bullet Cluster [8].

\begin{figure}
  \centering
  \includegraphics[width=0.25\textwidth]{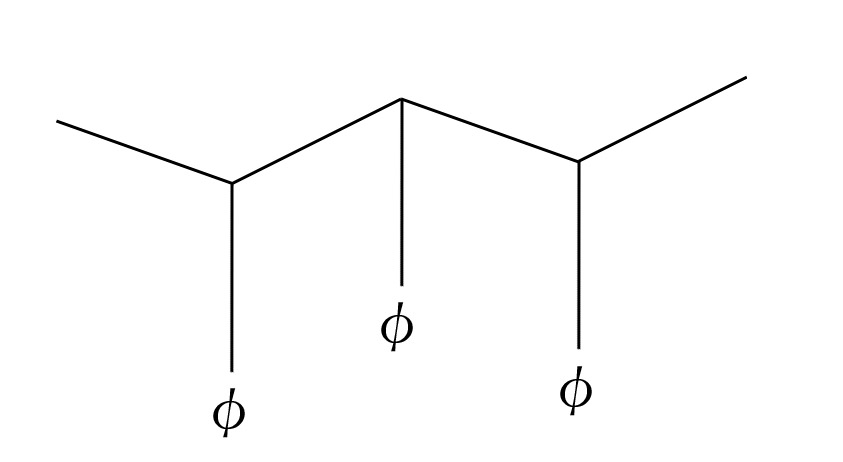}
  \caption{the dynamical particle hitting Higgs particles to gain the mass spans the mean free path.}\label{f3}
\end{figure}

\section{$\Psi$DM}
Adding dynamical term to $L_{mass}$, we have

\begin{flalign}
L_f=-\frac{1}{4}Y_{\mu v}Y^{\mu v}+\frac{1}{2}m^{2}Y_{\mu}Y^{\mu},
\end{flalign}

where $Y_{\mu v}=\partial_{\mu}Y_{v} - \partial_{v}Y_{\mu}$. To describe $\Psi$DM , we consider the nonrelativistic approach of $Y_{\mu}$ first, and one may derive the condensation by regarding it as the dilute boson gas via self-gravity. Follow Proca's method [22]. The Schrodinger equation can be passed from Klein-Gordon equation of $Y_{\mu}$ with the ansatz, $Y_{\mu}=e^{-imt}\Psi_{\mu}$ and elimination of $\Psi_0$. We have

\begin{flalign}
  i\frac{\partial}{\partial t}\Psi_i &= -\nabla^2\Psi_i \nonumber
\end{flalign}

Besides, the dilute limit comes from the negligible interaction between $Y_{\mu}$ particles (we will discuss the interaction later). Therefore, in a self-gravity particle clump, particles in the dilute boson gas can form a Bose-Einstein Condensation. The N interacting bosons [23] can be written down as [24]

\begin{flalign}
  i\frac{\partial}{\partial t}\Psi & =[-\frac{\nabla^{2}}{2m}-Gm^{2}\int\frac{|\Psi|^{2}}{|\vec{r}-\vec{r}'|}d\vec{r}']\Psi, \nonumber
\end{flalign}

where we always freeze the spin degrees of freedom in BEC and simply set $\Psi_i=\Psi$ [25]. Also, we regard it as a classical wavefunction in depicting the collective mode of particles in BEC.  This approach gives us the well-known Schrodinger-Poisson equation [24] specializing in describing the condensation of dark matter with ultralight particles [7]. In our deduction, $Y_{\mu}$ plausibly fits the criteria of $\Psi$DM via not only the mechanism but the reasonably ultra small $g^{'''}$. In addition, resembling $Y_{\mu}$ as $\Psi$DM benefits the further discussion of Bullet Cluster.



\section{Bullet Cluster}


In the case of BC, the accumulated data [10], as mentioned earlier, limit the interaction of DM-DM scattering, but open a wide opportunity that DM may interact with ordinary matters. In this letter, the interaction Lagrangian provides a clue to distinguish the two different modes of scattering, DM-matter and DM-DM. In fermion-gauge field coupled Lagrangian, one may have

\begin{flalign}
  L_{int} & =i\bar{R}\gamma^{\mu}D_{\mu}R+i\bar{L}\gamma^{\mu}D_{\mu}L
\end{flalign}

In our model, there will always be accompanied a $U_\phi(1)$ when making a $U_Y(1)$ transformation. Moreover, in the limit of small $g^{''}$, $Y_{\mu}=C_{\mu}$ and one may independently discuss $U_\phi(1)$. Therefore, narrowing down to the interaction involving $Y_{\mu}$ gives

\begin{flalign}
  L'_{int} & =-ig^{''}K\bar{f}\gamma^{\mu}Y_{\mu}f
\end{flalign}

where f represents quarks or leptons and k depends on weak hypercharge [26]. The combination of $L'_{int}$ and criteria of BC motives us two possible interactions, the Compton-like interaction in Figure 4 and $Y_{\mu}-Y_{\mu}$ interaction (similar to $\gamma-\gamma$ interaction) in Figure 5, where $Y_{\mu}$ plays the role of photon.

\begin{figure}[p]
  \centering
  \includegraphics[width=0.25\textwidth]{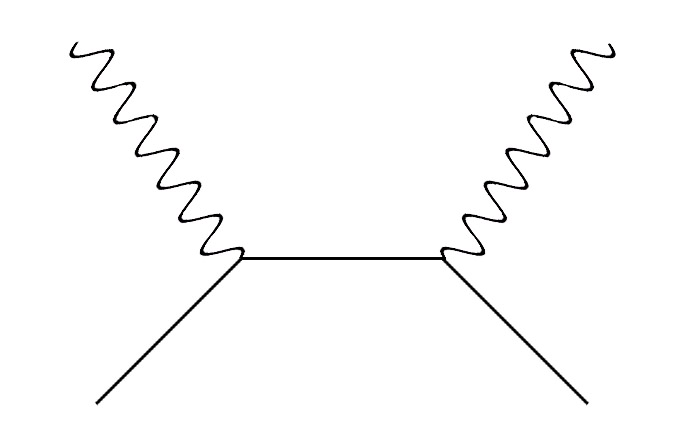}
  \caption{the scattering amplitude of this Compton-like scattering $\propto g^{''2}$  to the lowest order.}\label{f4}
\end{figure}

\begin{figure}[p]
  \centering
  \includegraphics[width=0.25\textwidth]{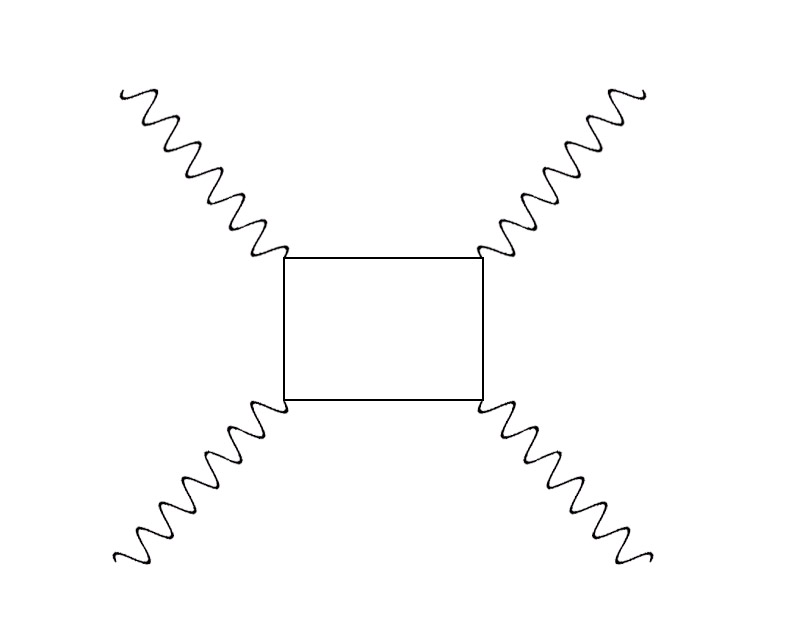}
  \caption{to mimic a $\gamma-\gamma$ interaction, the scattering amplitude of this $Y_{\mu}-Y_{\mu}$ scattering $\propto g^{''4}$ to the lowest order.}\label{f5}
\end{figure}

Although the coupling constant of scattering amplitude, $g^{''}$, is driven to a small value,  the  smallness of $g^{''2}$ may still be reflected by observing astrophysically in large particle collisions as in BC. On the other side, $g^{''4}$ is far below $g^{''2}$, this renders the amplitude of scattering  $Y_{\mu}-Y_{\mu}$, $\propto g^{''4}$, negligible. The assumption that DM particle is collisionless between each other still applies! Therefore, DM clumps may pass through each other unaffectedly but exert forces on ordinary particles.


\section{Prospect}

In the end, we may prospect the possible effect of coupling constant $g^{''}$, $g^{'''}$ and future work. Since the $\theta$ dependence in equation (2) is irrelevant to $\mu$, which implies the temperature independence, we may extrapolate $Y_{\mu}$ to interact with F in the early universe. Therefore, from the Lagrangian of (3), we expect an interacting process as shown in Figure 6.

\begin{figure}
  \centering
  \includegraphics[width=0.25\textwidth]{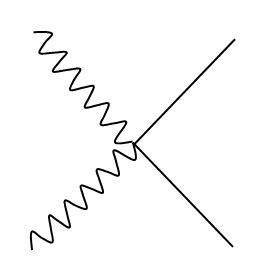}
  \caption{a vertex of $Y_{\mu}Y_{\mu}\rightarrow FF $ corresponds to a cross section $\propto g^{'''4}$ to the lowest order.}\label{f6}
\end{figure}

After reheating [16], when the inflatons decay into other particles and become dilute in the Universe, $g^{'''}$ drops gradually, which implies that $Y_{\mu}$ would ultimately decouple from the inflaton bath. But before that, a process called preheating phase [27] takes place in between inflation and reheating. In this period, particles and gauge fields [19], created and interacted by inflatons, will be resonantly amplified by coherently oscillating and causally produced explosively. We may expect an explosively increasing particle number of $Y_{\mu}$ in exponential growth (Appendix). In another aspect, if $g^{''}$ is comparable to $g$  and $g^{'}$ of electroweak couplings beyond the scale $\sim TeV$, an apparent decay of $Y_{\mu}\rightarrow f\bar{f}$ would largely decrease the abundance of $Y_{\mu}$ until $g^{''}$ drops far below $g$ and $g'$  to cease the interaction. Therefore the correct abundance of $Y_{\mu}$ should subtract the number of decayed  $Y_{\mu}$ from the one after reheating. By examining the detailed evolution of $g^{''}$ and $g^{'''}$, it would not only give the abundance but open opportunities to discuss the Standard Model in the vast energy gap from inflation phase to the pre-electroweak era.

\section{Acknowledge}

I would like to thank Yu-Hsiang Lin, Yen-Yu Lai, Jit-Liang Leong, Liang-Yao Wang, Chun-Fan Liu, Yan-Ting Chung and Fu-Goul Yee of National Taiwan University, who benefit me greatly from the discussion and especially express my gratitude to professor Tung-Mow Yan of Cornell University for the useful discussion and his kindly instruction.

\section{Appendix}

In preheating phase, we consider $S=\int d^{4}x\sqrt{-g}(\frac{R}{16\pi G}+L_{inf laton}-\frac{1}{4}(Y_{\mu v})^{2}+g^{'''2}(Y_{\mu}Y^{\mu})F^{2})$ .  Variation with respect to $Y^{\mu}$ leads to $\frac{1}{\sqrt{-g}}\frac{\partial}{\partial x^{\mu}}(\sqrt{-g}Y^{\mu v}) + 2g^{'''}F^{2}Y^{v}=0$. Choosing Robertson-Walker metric $ds^2=dt^2-a^2(t)d\vec{x}^2$ and $Y_0=0$ in Lorenz gauge, we have $\ddot{Y}_{i}+\frac{\dot{a}}{a}\dot{Y}_{i}-\frac{1}{a^{2}}\delta Y_{i}+2 g^{'''2}F^{2}Y_{i}=0$ Expanding $Y_i$ into Fourier space yields $\ddot{Y}_k+\omega_{k}^{2}Y_{k}=0$, where $\omega_{k}^{2}=k^{2}+2g^{'''2}f^{2}\sin^{2}mt$ after setting $a=0$, $\dot{a}=0$, $F=f\sin mt$ and $f\approx const$. [27]. The equation performs as a harmonic oscillator with time dependent $\omega_k$. Similar to lattice vibration, large particles bounded coherently will correspond to $E_k=N_k\omega_k$. We simply term $E_k=\frac{1}{2} \dot{Y}_k^2 + \frac{1}{2}\omega_k^2 Y_k^2$ in the approach. Moreover, by setting $z=mt$, $p=\frac{k^2}{m^2} +2q$ and $q=\frac{g^{'''2}f^2}{4m^2}$, we get the Mathieu-like equation $Y^{''}_k+(p-2q\cos (2z))Y_k=0$, where the prime denotes the derivative with respect to z. Solving this equation leads to Floguet solutions, $Y_k(z)=e^{i\mu_k z}P(p,q,z)$, in which P is a periodic function and $\mu_k$ is the exponent of complex number. The instability condition [27], when $i\mu_k=v_k>0$, shows an exponential magnification of $Y_k$. Therefore, the occupancy number $N_k=\frac{E_k}{\omega_k}\propto e^{2v_kmt}$, and $N_Y=\int \frac{d^3k}{(2\pi)^3 }N_k$, showing an exponential growth.






\section{References}

[1] V.C. Rubin, W.K. Ford, Jr. and N. Thonnard, Astrophys. J. 238 (1980) 471.

[2] J. R. Bond, A. S. Szalay and M. S. Turner, Phys. Rev. Lett. 48, 1636 (1982); G. R. Blumenthal, H. Pagels and J. R. Primack, Nature, 299, 37 (1982); J. Peebles, Astrophys. J. 263 (1982) L1 and G. R. Blumenthal, S. M. Faber, J. R. Primack and M. J. Rees, Nature 311, 517 (1984).

[3] W. J. G. de Blok, Advances in Astronomy, article id. 789293 (2010).

[4] J. Dubinski and R. G. Carlberg, Astrophys. J. 378, 496 (1991); J. F. Navarro, C. S. Frenk, and S. D. M. White, Astrophys. J. 462, 563 (1996); J. F. Navarro, C. S. Frenk and S. D. M. White, Astrophys. J. 490, 493 (1997).

[5] B. Moore, 1994, Nature 370, 629; R.A. Flores and J.R. Primack, Astrophys. J. 427, L1 (1994).

[6] W. J. G. de Blok and A. Bosma, Astron. Astrophys. 385, 816 (2002).

[7] W. Hu, R. Barkana, and A. Gruzinov, Phys. Rev. Lett. 85, 1158 (2000); P. J. E. Peebles, Astrophys. J. 534, L127 (2000).

[8] D. Clowe, M. Bradac, A. H. Gonzalez, M. Markevitch, S. W. Randell, C. Jones and D. Zaritsky, Astrophys. J. 648, L109 (2006).

[9] H.-Y. Schive, T. Chiueh, and T. Broadhurst, Nat. Phys. 10, 496 (2014).

[10] D. Harvey, R. Massey, T. Kitching, A. Taylor, and E. Tittley, Science 347, 1462 (2015).

[11] Axion has been considered as the candidate of $\Psi$DM; however, the energy scale itself is incompatible with $\Psi$DM, as pointed out by L. M. Widrow Nat. Phys. 10, 477 (2014). Besides, the dynamics of axion seems difficult to explain the Bullet Cluster phenomena.

[12] We do not consider the mechanism of Modified Newtonian Dynamics (MOND). See e.g. G. W. Angus, B. Famaey, and H. Zhao, S. 2006, MNRAS, 371, 138. We wish to give a perspective from particle physics. The criticisms of MOND can be seen from A. Aguirre, J. Schaye, and E. Quataert, Astrophys. J. 561, 550 (2001); D. Scott, M. White, J. D. Cohn and E. Pierpaoli, arXiv:astro-ph/0104435. Since the cross section of the self-interaction between DM particles is limited from the BC observation, we also exclude it from our discussion. See M. Markevitch, A. H. Gonzalez, D. Clowe, A. Vikhlinin,  W. Forman, C. Jones, S. Murray and W. Tucker, Astrophys. J. 606, 819 (2004).

[13] A. H. Guth, Phys. Rev. D 23, 347 (1981); A. D. Linde, Phys. Lett. B 108, 389 (1982); A. Albrecht and P. J. Steinhardt, Phys. Rev. Lett. 48, 1220 (1982).

[14] G. F. Smoot et al., 1992, ApJ, 396, L1; C. L. Bennett et al., Astrophys. J. Suppl. Ser. 208 (2013) 20; P. A. R. Ade et al., Astron. Astrophys. 571, A1 (2014); P. A. R. Ade et al., arXiv:1502.01589.

[15] D. Baumann, TASI lectures on inflation, arXiv:0907.5424.

[16] L. Kofman, A.D. Linde, A.A. Starobinsky, Phys. Rev. Lett. 73, 3195 (1994); Y. Shtanov, J.H. Traschen, R.H. Brandenberger, Phys. Rev. D 51, 5438 (1995); L. Kofman, A.D. Linde, A.A. Starobinsky, Phys. Rev. D 56, 3258 (1997).

[17] M. E. Peskin and D. V. Schroeder, An Introduction to Quantum Field Theory, Westview Press, Boulder, CO, 1995.

[18] L. M. Widrow and N. Kaiser, Astrophys. J. 416, L71 (1993).

[19] B. A. Bassett, G. Pollifrone, S. Tsujikawa and F. Viniegra, Phys. Rev. D 63, 103515 (2001); F. Finelli and A. Gruppuso, Phys. Lett. B 502, 216 (2001); A. Rajantie and E. J. Copeland, Phys. Rev. Lett. 85, 916 (2000).

[20] E. N. Parker, Cosmical Magnetic Fields (Clarendon, Oxford, England, 1979); Y. Zel'dovich, A. Ruzmaikin, and D. Sokoloff, Magnetic Fields in Astrophysics (Gordon and Breach, New York, 1983).

[21] To estimate the relation of $g^{''}$ and coupling c, in a classical picture we set $\theta=\theta_0+\delta$ and expand around $\theta_0$  as shown in Fig. 2, such that $ \Delta V =\Delta E =V(\theta)-V(\theta_0)=
\frac{\delta^2c}{2}(a\cos (\theta_0)+ b\sin (\theta))$. If the perturbation $\Delta E$  converts energy to $Y_{\mu}$ to accumulate mass energy, we naively expect the simple relation $g^{'''}\propto c$.

[22] A. Proca, J. Phys. Rad. 9 (1938), 61-66.

[23] E. P. Gross, 1961, Nuovo Cimento 20, 454; L. P. Pitaevskii, 1961, Sov. Phys. JETP 13, 451.

[24] E.J.M. Madarassy and V.T. Toth, Phys. Rev. D 91, 044041 (2015).

[25] The spin degrees of freedom can be seen from W. Greiner and J. Reinhardt, Field Quantization, Springer, 1995, while the spin contribution in BEC can be seen from T. Ohmi and K. Machida, J. Phys. Soc. Jpn. 67, 1822 (1998). However, in our model the driving field is gravitation, we may ignore the spin effect.

[26] As in the case of electroweak theory, there is also a $j^3_\alpha $, the third component of isovector current.

[27] R. Allahverdi, R. Brandenberger, F.-Y. Cyr-Racine and A. Mazumdar, 2010, Ann. Rev. Nucl. Part. Sci. 60 27; L. Kofman, arXiv:hep-ph/9802285.




\end{document}